\documentclass[nohyper,12pt,letterpaper]{JHEP}
\usepackage[dvips]{epsfig}
\usepackage{epsf,amsfonts,amssymb}
\usepackage{graphicx}
\usepackage{amsmath}

\input epsf.sty
\input{tcilatex}

\title{The Decay of The Five Brane\\
in $AdS_5\times \mathbb{R}P^5$\footnote{Lecturenotes given by the author at the Physics Department
of the Turin University on October 2002}}
\author{El Mostapha Sahraoui \\
National Grouping of High Energy Physics, GNPHE.\\
\qquad\qquad\qquad\qquad\qquad and\\
Lab/UFR  High Energy Physics,\\
Physics Department, Faculty of
Sciences, P.O. Box 1014, \\Av Inb Batouta, Rabat, Kingdom of Morocco.\\
E-mail:
\email{mostafa sahraoui@netscape.net, lphe@fsr.ac.ma} }

\abstract{The baryon vertex of IIB superstring theory on
$AdS_{5}\times \mathbb{R}P^{5}$, for the case of orthogonal
groups, is studied. The energy of the three brane decayed from an
original five brane is calculated explicitly. The radius of this
decayed three brane, for a BPS configuration, is also given and
interpreted.}

\preprint{ hep-th/\\ GNHEP/03/03} 
\begin{document}
\section{Introduction}

The conjecture of Maldacena\ \cite{malda}\ says that the
four-dimensional $\mathcal{N}=4$ super Yang Mills theory with
gauge group $SU(N)$ is equivalent to Type IIB superstring theory
on $AdS_{5}\times \mathbf{S}^{5}$ where $AdS_{5}$ is the
five-dimensional Anti de Sitter space and $N$ the five-form flux
on the five sphere representing the number of the parallel D3
branes on which the theory lives. Soon after this important
discovery, several applications to systems with reduced
supersymmetry, that are obtained by an orbifolding operation in
which $\mathbf{S}^{5}$ was replaced by $\mathbf{S} ^{5}/\Gamma $
with $\Gamma $\ a finite group, was investigated \cite
{KacSil,AOY,Kak,Kak2,FaySpa}.

In particular interest, Witten in \cite{witten} has studied the
case of the orientifolding operation in which $\mathbf{S}^{5}$ is
replaced by $\mathbb{R} P^{5}=\mathbf{S}^{5}/\mathbb{Z}_{2}$ and
where the gauge group $SU(N)$ is now replaced by an orthogonal
$SO(N)$ or symplictic $Sp(N)$ gauge groups. Moreover, the author
has discussed the possibility of wrapping branes depending in the
discrete torsion and has given interpretations to known examples
of the gauge theory in terms of branes in $\mathbb{R}P^{5}$ as
pfaffian, domain walls and baryon vertex.

In $AdS_{5}\times \mathbf{S}^{5}$, the baryon configuration consists of five
brane wrapped around $\mathbf{S}^{5}$ and joined to the boundary by $N$
fundamental strings\ with the same orientation \cite{imamura2,BISY,CGS}. The
wrapped five brane feels two forces: the gravitational force caused by the $N
$ D3 branes and the tension of $N$ fundamental strings attached on it. On
the other hand, in a supersymmetric configuration, there is an effect of the
electric field living on the brane to be not neglected and that can deform
the brane. Therefore, the correct energy of the configuration is obtained by
using the Born-Infeld action of D-branes \cite{gibbons,CM}. Based on the
previous works, authors in\ \cite{imamura} were able\ to compute the bending
energy and the radius of this baryon vertex starting from its BI action in $%
AdS_{5}\times \mathbf{S}^{5}$, see also \cite{CGMvP,GRST}.

In the present paper, we generalize this analysis by studying the
decay of the five brane in $AdS_{5}\times \mathbb{R}P^{5}$ in the
case of the orthogonal group leading to a three brane
configuration attached by a string or a three brane configuration
alone. We compute explicitly the energy of this configurations and
give the expression of the decayed three brane energy and radius
in terms of the collatitude angle. We also give an interpretation
of the result based on diagram ullistrations.

This paper is organized as follows. In section 2, we give a brief
review on baryon vertex in terms of branes. In section 3, we
discuss the possibility of wrapping branes of type IIB superstring
on $AdS_{5}\times \mathbb{R}P^{5}$ for symplectic and orthogonal
groups. In section 4, we focus on five brane and its decay to
three brane in the case of orthogonal group. In section 5, we
compute the energy of the decayed three brane configuration and
give the expression of its radius. The final section is devoted to
the conclusion


\section{Baryonic D5-Brane}

Let us here summarize in few lines what does a baryonic D5-Brane
mean. Following the correspondence between IIB on $AdS_{5}\times
\mathbf{S}^{5}$ space and conformal field theory on its boundary,
one would like to find, in
terms of string theory, the equivalent of a static baryon vertex of the $%
SU(N)$ gauge theory namely a static gauge invariant antisymmetric
combination of $N$ external electric quarks. In fact, it consists
of a static wrapped D5-brane of type IIB superstring theory
centered in the bulk and joined to the boundary by $N$ elementary
strings. In this configuration, the static external quarks in the
SYM theory, that is static external electric charges transforming
in the fundamental representation of $SU(N)$, are described by the
endpoints of the fundamental strings (see figure \ref{a}). What
plays the role of the ``glue'' for this vertex is the coupling

\FIGURE[t]{
\epsfig{file=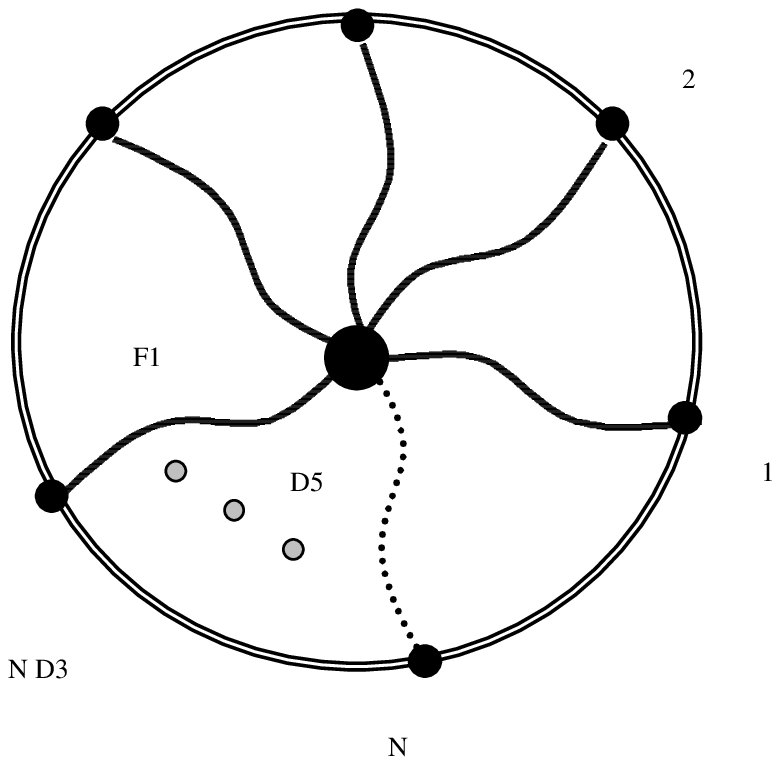, height=8cm}
\caption{Baryon vertex of IIB superstring on $AdS_5\times \mathbf{S}^{5}$
representing by a wrapped five brane on $\mathbf{S}^{5}$ and attached to the boundary
by $N$ strings.}
\label{a}
}

\begin{equation}
\int_{\mathbf{S}^{5}\times \mathbf{R}}A\wedge \frac{G_{5}}{2\pi }
\end{equation}
between the self-dual five form field strength $G_{5}$, contributing with $N$
units of five form flux on $\mathbf{S}^{5}$, and the $U(N)$ gauge field $A$
living on the world volume of the D5-brane.

Moreover, the deformation of the shape of the D5-brane caused by the
tension of the strings is given by the radial position $r$ of the D5-brane in $%
AdS$ space as a function of the collatitude angle $\theta $\ as
\begin{equation}
r\left( \theta \right) \sim \frac{\left( \pi -\theta +\sin \theta \cos
\theta \right) ^{1/3}}{\sin \theta }  \label{shape}
\end{equation}
we see from eq (\ref{shape})\ that $r\left( \theta \right) $\
diverges for small $\theta $'s representing the $N$ fundamental
strings connecting the D5 brane to the boundary\ at $r\rightarrow
\infty $.


\section{Branes on $\mathbb{R}P^{5}$}

In this section we select the whole possibilities of wrapping branes on $\mathbb{R}%
P^{5} $. But before going ahead, recall that the $SO(N)$ (for even $N$) and $%
Sp(N)$ gauge symmetries can be obtained by considering $N$ parallel threebrane
at an orientifold threeplane. In contrast to the $SU(N)$ gauge theory case,
where the near horizon geometry of $N$ parallel threebranes in $\mathbb{R}^{10}$ lead to $%
AdS_{5}\times \mathbf{S}^{5}$ description, here and under $\mathbb{Z}_{2}$ action on $%
\mathbb{R}^{6}$ part of $\mathbb{R}^{10}$, one gets in the near horizon
geometry $AdS_{5}\times \mathbb{R}P^{5}$.

Thus, the $\mathcal{N}=4$ super Yang-Mills theory with orthogonal or
symplectic gauge group can be described by Type IIB superstring theory on an
$AdS_{5}\times \mathbb{R}P^{5}$ orientifold \cite{AOY, Kak, Kak2, FaySpa}.
The spectra of the $SO(N)$ and $Sp(N)$ gauge theory can be
obtained from those of the $SU(N)$ theory by extracting the part invariant
under an orientifold projection.\newline

\bigskip
\noindent
\textit{Homology and Cohomology of $\mathbb{R}P^{5}$}\newline

In type IIB superstring theory, there exist a supersymmetric
orientifold threeplane that is invariant under the
$SL(2,\mathbb{Z})$ S-duality symmetry group leading to an
$SL(2,\mathbb{Z})$ invariant configuration of threebranes on
$AdS_{5}\times \mathbf{S}^{5}/ \mathbb{Z}_{2}$ (after taking the
near horizon geometry of $\mathbb{R}^{4}\times\mathbb{R}^{6}
/\mathbb{Z}_{2}$). In addition to this, the two two-form fields:
the Neveu Schwarz $B$ field $B_{NS}$ and Ramond-Ramond $B$ field
$B_{RR}$ define four other models determined by the values
 of their discrete torsions.

The different homology groups that will play a central role in the
wrapping branes are summarized here after. For odd $i$,
$\mathbb{R}P^{i}$ determines an element of
$H^{i}(\mathbb{R}P{^{5}},\mathbb{Z})$ and if $i$ is even it determines an
element of $H^{i}(\mathbb{R}P{^{5}},\widetilde{\mathbb{Z}})$, where $\mathbb{R}P^{i}$
is a subspace of $\mathbb{R}P^{5}$ defined by a linear embedding $%
(x_{1},x_{2},\ldots ,x_{i+1})\rightarrow (x_{1},x_{2},\ldots
,x_{i+1},0,\ldots ,0)$. The homology groups generated by the two torsion
element defined by these subspaces are
\begin{eqnarray}
H_{1}(\mathbb{R}P^{5},\mathbb{Z}) &=&H_{3}(\mathbb{R}P^{5},\mathbb{Z})=%
\mathbb{Z}_{2} \\
H_{0}(\mathbb{R}P^{5},\mathbb{Z}) &=&H_{5}(\mathbb{R}P^{5},\mathbb{Z})=%
\mathbb{Z}
\end{eqnarray}
and
\begin{eqnarray}
H_{2}(\mathbb{R}P^{5},\widetilde{\mathbb{Z}}) &=&H_{4}(\mathbb{R}P^{5},%
\widetilde{\mathbb{Z}})=\mathbb{Z}_{2} \\
H_{0}(\mathbb{R}P^{5},\widetilde{\mathbb{Z}}) &=&\mathbb{Z}_{2},
\end{eqnarray}
where $\widetilde{\mathbb{Z}}$ is the twisted sheaf of integers.
The Poincar\'{e} duality permit us to get cohomology groups from
homology ones, thus we have
\begin{eqnarray}
H_{i}(\mathbb{R}P^{5},\mathbb{Z}) &=&H^{5-i}(\mathbb{R}P^{5},\mathbb{Z}) \\
H_{i}(\mathbb{R}P^{5},\widetilde{\mathbb{Z}}) &=&H_{5-i}(\mathbb{R}P^{5},%
\widetilde{\mathbb{Z}})
\end{eqnarray}

\bigskip \noindent \textit{Wrapping Branes in $\mathbb{R}P^5$}\newline

Return now to the possibilities of wrapping branes. At first site, a five
brane can be wrapped on a tow-cycle to give, in $AdS_{5}$, a threebrane or
on a four-cycle to give a string as $H_{2}(\mathbb{R}P^{5},\widetilde{%
\mathbb{Z}})=\mathbb{Z}_{2}$ for the former and $H_{4}(\mathbb{R}P^{5},%
\widetilde{\mathbb{Z}})=\mathbb{Z}_{2}$ for the later.

Similarly for the three brane, it can be wrapped on a one-cycle to give a two
brane or on three-cycle to give a particle on $AdS_{5}$ as here also $H_{1}(%
\mathbb{R}P^{5},\mathbb{Z})=\mathbb{Z}_{2}$ and $H_{3}(\mathbb{R}P^{5},%
\mathbb{Z})=\mathbb{Z}_{2}$

But this is not the end of the story as there is some topological
restriction based on discrete torsion. In fact, and as was
explained in \cite{witten},
 a D5-brane (NS5-brane) can be wrapped on an $\mathbb{R}P^{4}\subset \mathbb{R}P^{5}$,
to make a string, only if $\theta _{NS}=0$ ($\theta _{RR}=0$). And the three brane can be
wrapped on an $\mathbb{R}P{^{3}\subset }\mathbb{R}P^{5}$, to make a
particle, only if $\theta _{NS}=\theta _{RR}=0$.

In the other hand, the existing four models of gauge theories,
depending on whether the discrete torsion vanishes or not, are
classified in the following way as:
\begin{eqnarray}
(\theta _{NS}=0,\;\;\theta _{RR}=0)\qquad &\rightarrow &\qquad SO(N)\qquad {%
\ }\mathrm{for\;even}\;N \\
(\theta _{NS}=0,\;\;\theta _{RR}\neq 0)\qquad &\rightarrow &\qquad
SO(N)\qquad {\ }\mathrm{for\;odd}\;N \\
(\theta _{NS}\neq 0,\;\;\theta _{RR}=0)\qquad &\rightarrow &\qquad Sp(N) \\
(\theta _{NS}\neq 0,\;\;\theta _{RR}\neq 0)\qquad &\rightarrow &\qquad Sp(N).
\end{eqnarray}

Finally, a wrapping D5-brane on $\mathbb{R}P^{4}$ gives rise to orthogonal
gauge group, likewise an so wrapped NS5-brane gives rise to either $SO(N)$
for $N=2k$ or symplectic gauge group. While only the orthogonal group for $%
N=2k$ is permitted in the case of wrapping threebrane on $\mathbb{R}P^{3}$.

We conclude this section by saying that a five brane can be
wrapped around a two cycle in $\mathbb{R}P^{5}$ to give a three
brane with orthogonal gauge group.


\section{Baryon Vertex for Orthogonal and Symplectic Groups}

Let us summarize here the stability of the baryonic D5-brane in $\mathbb{R}%
P^{5}$ depending on the nature of the gauge groups \cite{witten}. For the
case of the symplectic gauge group i.e. $\theta _{NS}\neq 0$, there exist
always mesons $M=\frac{1}{2}\gamma _{ij}\psi ^{i}\psi ^{j}$ to which a
baryon $B=\frac{1}{N!}\varepsilon _{i_{1}i_{2}...i_{N}}\psi ^{i_{1}}\psi
^{i_{2}}...\psi ^{i_{N}}$ can decay as
\begin{equation}
B=\frac{1}{(N/2)!}M^{N/2}
\end{equation}
where $\gamma _{ij}$ is an invariant second rank antisymmetric tensor and $%
\psi $ a fermion in the fundamental representation of the gauge group.
Therefore, an initial state with a fivebrane wrapped twice on $\mathbb{R}%
P^{5}$ and connected by $N$ elementary strings to charges on the boundary is
able to decay to a state with no fivebrane and with $N/2$ strings that join
the external quarks pairewise.
\FIGURE[t]{
\epsfig{file=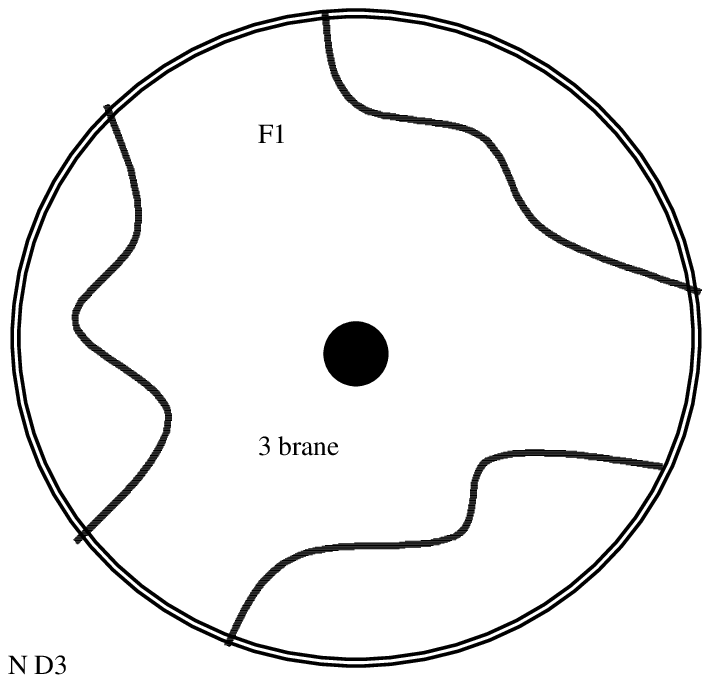, height=8cm}
\caption{The final state of the decay of a five brane in $AdS_5\times \mathbb{R}P^{5}$ for
$SO(N)$ gauge group with $N$ even. It is a three brane wrapped on a three cycle in $\mathbb{R}P^{5}$
 and strings joined to the boundary pairewise.}
\label{b}
}

In the case of the orthogonal gauge group that is for
$\theta_{NS}= 0$, we have to distinguish between $N$ even and odd.
In the former case, the final state is a Pfaffian combination of
$N/2$ gauge bosons interpreted as a wrapped threebrane plus
strings making paierwise connections between external charges (see
figure \ref{b}). While in the later, it contains in addition an
odd number of strings connecting the wrapped threebrane to the
boundary (see figure \ref{c}).
\FIGURE[t]{ \epsfig{file=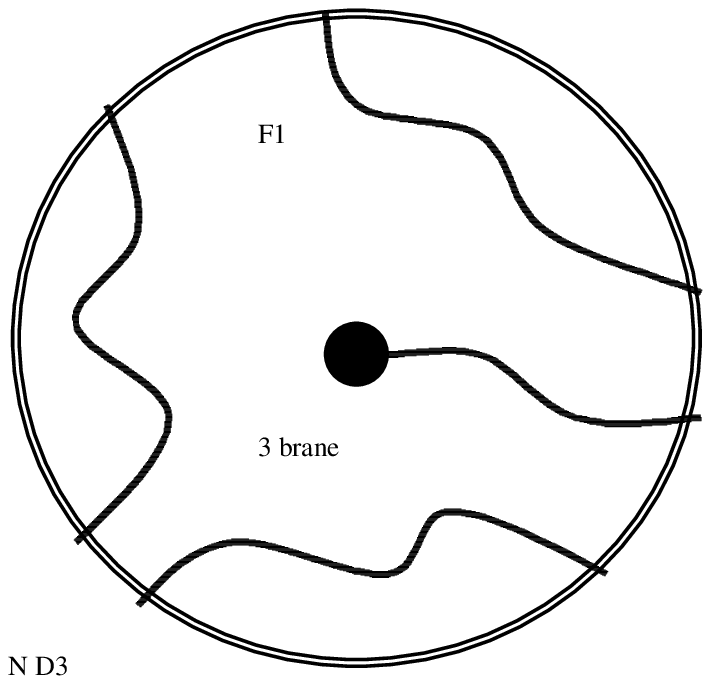, height=8cm} \caption{This
final states represents the decay of a five brane in $AdS_5\times
\mathbb{R}P^{5}$ for $SO(N)$ gauge group with $N$ odd. It is a
three brane wrapped on a three cycle in $\mathbb{R}P^{5}$ and a
string attached on it in addition to strings joined to the
boundary pairewise.} \label{c} }


\section{Decay of D5-Brane}

In this section we concentrate our attention to the decay of the D5-brane to
three brane i.e. the case of
orthogonal gauge group corresponding to the vanishing value of $%
\theta _{NS}$. We will give the equivalent, in our case, of the eq
(\ref {shape}), expressing the behavior of the radius of the
D5-brane in terms of the collatitude angle $\theta $, and discuss
its behavior at extreme limits. To do so, one may start directly
from the metric of a D3 brane but this will be not interesting in
our study as we plan to describe the decay of the baryonic
D5-brane in $AdS_{5}\times \mathbb{R}P^{5}$ to a three brane of
type IIB super string theory. Our philosophy is to start from the
action of the D5-brane and then we arrange to express it in terms
of three brane one. Indeed, one way to write a D5-brane metric in
the background geometry of a stack of $N$ D3-branes is the
following
\begin{equation}
ds^{2}=\frac{r^{2}}{R^{2}}\left[ -dt^{2}+dx_{||}^{2}\right] +\frac{R^{2}}{%
r^{2}}\left[ dr^{2}+r^{2}d\Omega _{5}^{2}\right]  \label{d5metric}
\end{equation}
where $x_{||}$\ are the three dimensional Euclidean $\mathbb{E}^{3}$
coordinates and $d\Omega _{5}^{2}$\ is the line element on the five sphere $%
S^{5}$. Then the world volume action of this D5-brane is the Born-Infeld
action given by
\begin{equation}
\mathcal{S}=-T_{5}\int d^{6}\zeta \sqrt{-\det (g^{\mathrm{ind}})}+T_{5}\int
d^{6}\zeta A_{\alpha }\partial _{\beta }X^{M_{1}}\wedge ...\wedge \partial
_{\gamma }X^{M_{5}}G_{M_{1}..M_{5}}  \label{d5action}
\end{equation}
with
\begin{equation*}
g_{\alpha \beta }^{\mathrm{ind}}=g_{MN}\partial _{\alpha }X^{M}\partial
_{\beta }X^{N}+\mathcal{F}_{\alpha \beta }
\end{equation*}
and where $T_{5}$\ is the D5 brane tension and the second term in eq (\ref
{d5action}) is the explicit WZW coupling of the world volume gauge field $A$
to the background five form field strength $G$.

Now, given this metric at hand, eq (\ref{d5metric}), we can decompose its
last term as
\begin{equation*}
ds^{2}=\frac{r^{2}}{R^{2}}\left[ -dt^{2}+dx_{||}^{2}\right] +\frac{R^{2}}{%
r^{2}}\left[ dr^{2}+r^{2}\left( d\theta ^{2}+\cos ^{2}\theta d\psi ^{2}+\sin
^{2}\theta d\Omega _{3}^{2}\right) \right]
\end{equation*}
with
\begin{equation*}
d\Omega _{3}^{2}=d\theta ^{\prime 2}+\cos ^{2}\theta ^{\prime }d\psi
^{\prime 2}+\sin ^{2}\theta ^{\prime }d\phi
\end{equation*}
describing a line element of the three sphere $S^{3}$.

Now our strategy is to choose the world volume coordinates for the
D5-brane as
\begin{equation*}
\zeta _{\alpha }=\left( t,\theta ,\psi ,\theta ^{\prime },\psi ^{\prime
},\phi \right)
\end{equation*}
where $t$ is the target space time, and then set the space time coordinates
as
\begin{equation*}
X^{M}=\left( t,x_{||},r,\theta ,\psi ,\theta ^{\prime },\psi ^{\prime },\phi
\right)
\end{equation*}
Furthermore, we suppose that the radius of the decayed threebrane
is described by $\mathfrak{r}$\ such that
\begin{equation}
\frak{r}=r\sin \theta .
\end{equation}
The key idea is to fix $\theta $\ and look for a static solutions
of the form $\frak{r}\left( \theta ^{\prime }\right) $ and
$A_{0}\left( \theta ^{\prime }\right) $, with the $\theta ^{\prime
}$\ is interpreted as representing an angle from the opposite
point to the string endpoint. The other four parameters $\theta
,\psi ,\psi ^{\prime }$and $\phi $ are angular
variables parameterizing $\mathbb{R}P^{4}\subset \mathbb{R}P^{5}$ with fixed $%
\theta ^{\prime }$. Thus, the only two independent variables are
$\frak{r}$ and $\theta ^{\prime }$, so
\begin{equation*}
\frak{r}^{\prime }=\frac{\partial \frak{r}}{\partial \theta ^{\prime }}=%
\frac{\partial }{\partial \theta ^{\prime }}\left( r\sin \theta \right)
=r^{\prime }\sin \theta ,
\end{equation*}
and
\begin{equation*}
{\dot{\frak{r}}}=\frac{\partial \frak{r}}{\partial t}=\frac{\partial }{%
\partial t}\left( r\sin \theta \right) =\dot{r}\sin \theta .
\end{equation*}


\section{Three brane energy}

All materials at hand, we can now compute the binding energy of
the decayed three brane and its radius. But before going ahead let
us rewrite the metric
following the assumption of the previous section as a metric on $%
AdS_{2}\times \mathbb{R}P^{3}$
\begin{equation*}
ds^{2}=-\frac{\frak{r}^{2}}{\sin ^{2}\theta R^{2}}dt^{2}+\frac{\sin
^{2}\theta R^{2}}{\frak{r}^{2}}\left[ dr^{2}+\frak{r}^{2}d\theta ^{\prime 2}+%
\frak{r}^{2}\sin ^{2}\theta ^{\prime }d\phi +\frak{r}^{2}\cos ^{2}\theta
^{\prime }d\psi ^{\prime 2}\right]
\end{equation*}
Then the induced metric on the three brane is given by
\begin{equation*}
h_{\alpha \beta }=\left[
\begin{array}{cc}
-\frac{\frak{r}^{2}}{\sin ^{2}\theta R^{2}} & F_{0\theta ^{\prime }} \\
-F_{0\theta ^{\prime }} & \frac{R^{2}}{\frak{r}^{2}}\left[ \frak{r}^{\prime
2}+\frak{r}^{2}\sin ^{2}\theta \right]
\end{array}
\right]
\end{equation*}
whose determinant is
\begin{equation*}
\det h_{\alpha \beta }=-\left( \frac{\frak{r}^{\prime 2}}{\sin ^{2}\theta }+%
\frak{r}^{2}\right) +F_{0\theta ^{\prime }}^{2}.
\end{equation*}
Finally, one can derive the three brane energy starting from
D5-brane one as
\begin{equation}
\frak{E}=\mathrm{T}\int \mathrm{d}\theta \mathrm{d}\theta ^{\prime }\sin
^{4}\theta \left\{ -R^{4}\sqrt{\frac{\frak{r}^{\prime 2}}{\sin ^{2}\theta }+%
\frak{r}^{2}-F_{0\theta ^{\prime }}^{2}}+4A_{0}R^{4}\right\}  \label{energy}
\end{equation}
with $\mathrm{T}=T_{5}\Omega _{3}$,\ where $\Omega _{3}$\ denote
the unit three sphere.

\FIGURE[t]{ \epsfig{file=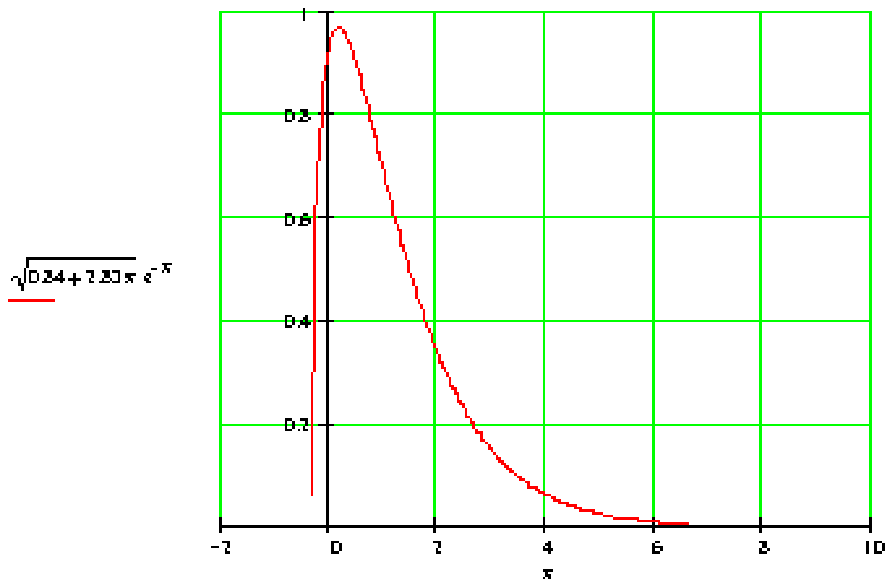, height=10cm} \caption{The
shape of the three brane radius for fixed $\theta=\pi/4$.}
\label{d} }

Then, the gauge field equation of motion following from this energy reads as
\begin{equation}
\partial _{\theta ^{\prime }}\left[ \frac{E^{\prime }}{\sqrt{\frac{\frak{r}%
^{\prime 2}}{\sin ^{2}\theta }+\frak{r}^{2}-E^{\prime }}}\right] =-4,
\label{eqmvt}
\end{equation}
with $F_{0\theta ^{\prime }}=E^{\prime }$ and if we interpret the term
between parenthesis as the three brane displacement
\begin{equation}
\mathcal{D}\left( \theta ,\theta ^{\prime }\right) =\frac{E^{\prime }}{\sqrt{%
\frac{\frak{r}^{\prime 2}}{\sin ^{2}\theta }+\frak{r}^{2}-E^{\prime }}},
\label{displacement}
\end{equation}
then eq(\ref{eqmvt}) becomes
\begin{equation}
\partial _{\theta ^{\prime }}\mathcal{D}\left( \theta ,\theta ^{\prime
}\right) =-4.  \label{eqmvtf}
\end{equation}

To resolve eq (\ref{eqmvtf}), we come back to our starting
assumption, that a D5 brane decay to a three brane, so we argue
that this differential equation should be treated in taking in the
account this previous detail. Thus, we propose that its solution
should be given by
\begin{equation}
\mathcal{D}\left( \theta ,\theta ^{\prime }\right) =-4\theta ^{\prime
}+D\left( \theta \right)  \label{solutree}
\end{equation}
where $D\left( \theta \right) $\ is the displacement of the original D5
brane given by
\begin{equation}
D\left( \theta \right) =-\frac{3}{2}\theta +\frac{3}{2}\sin \theta \cos
\theta +\sin ^{3}\theta \cos \theta .  \label{solufive}
\end{equation}
It is more useful to express the energy eq (\ref{energy}) in terms of $%
\mathcal{D}$\ so
\begin{equation*}
\frak{E}=\mathrm{T}\int \mathrm{d}\theta \mathrm{d}\theta ^{\prime }\sin
^{4}\theta R^{4}\left\{ \sqrt{\frac{\frak{r}^{\prime 2}}{\sin ^{2}\theta }+%
\frak{r}^{2}-E^{\prime 2}}+\mathcal{D}E^{\prime }\right\}
\end{equation*}
The final form of the energy is given after the elimination of $E^{\prime }$%
\ in terms of $\mathcal{D}$ using eq (\ref{displacement}) and
\begin{equation*}
\sqrt{\frac{\frak{r}^{\prime 2}}{\sin ^{2}\theta }+\frak{r}^{2}-E^{\prime 2}}%
=\frac{\sqrt{\frac{\frak{r}^{\prime 2}}{\sin ^{2}\theta }+\frak{r}^{2}}}{%
\sqrt{\mathcal{D}^{2}+1}},
\end{equation*}
so we get by the end the expression of decayed three energy brane as a
function of its displacement and radius
\begin{equation}
\frak{E}=\mathrm{T}\int \mathrm{d}\theta \mathrm{d}\theta ^{\prime }\sin
^{4}\theta R^{4}\sqrt{\mathcal{D}^{2}+1}\sqrt{\frac{\frak{r}^{\prime 2}}{%
\sin ^{2}\theta }+\frak{r}^{2}}.
\end{equation}

Now we are ready to compute the Euler Lagrange equations and
deduce the differential equation of the radius. Indeed a
straightforward calculation lead to
\begin{equation}
\frac{\mathrm{d}}{\mathrm{d}\theta ^{\prime }}\left\{ \frac{\frak{r}^{\prime
}}{\sqrt{\frac{\frak{r}^{\prime 2}}{\sin ^{2}\theta }+\frak{r}^{2}}}\sqrt{%
\mathcal{D}^{2}+1}\right\} =\frac{\frak{r}\sin ^{2}\theta }{\sqrt{\frac{%
\frak{r}^{\prime 2}}{\sin ^{2}\theta }+\frak{r}^{2}}}\sqrt{\mathcal{D}^{2}+1}%
,  \label{eqdiff}
\end{equation}
from wich one can extract the desired expression of the radius. But as we
are looking for a BPS solution, we argue that a the equation (\ref{eqdiff})
is reduces to
\begin{equation}
\frac{\frak{r}^{\prime }}{\frak{r}}=\frac{\sin \theta +\mathcal{D}\cos
\theta }{\cos \theta -\mathcal{D}\sin \theta }  \label{eqdiffparticulare}
\end{equation}
whose solution can be given as
\begin{equation}
\frak{r}=\left( \left[ \cos \theta -D\sin \theta \right] +4\theta ^{\prime
}\sin \theta \right) ^{\frac{1}{4}\left( 1+\alpha ^{2}\right) }.\exp \left(
-\alpha \theta ^{\prime }\right)  \label{finalesolut}
\end{equation}
where $\alpha =\frac{1}{\tan \theta }$ and $D$ is given by eq (\ref{solufive}%
).

\bigskip
\noindent \emph{Discussion}
\FIGURE[t]{
\epsfig{file=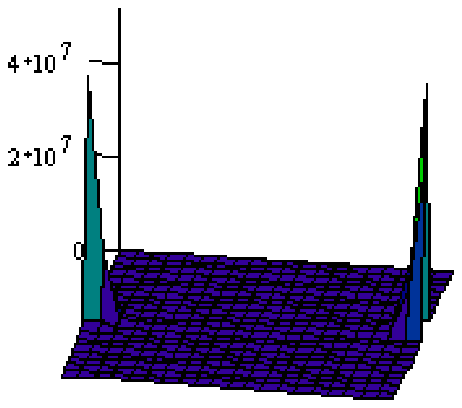, height=12cm}
\caption{The behavior of the radius in terms of ${\theta}^{\prime}$ and small $\theta$
.}
\label{e}
}

Let us now interpret the solution of the radius eq
(\ref{finalesolut}) in terms of $\theta ^{\prime }$ and $\theta $.
It is clear, due to the exponential term, that the expression
eq(\ref{finalesolut}) approaches zero. For a special value of
$\theta =\frac{\pi }{4}$, we see that the radius rich a maximum
than decrease rapidly  (see figure \ref{d}) the same behavior can
be remarked for small values of $\theta $ (figure \ref{e}). One
can comment this by saying that the three brane radius rich its
maximum for values of $\theta ^{\prime }$\ very close to zero then
once $\theta ^{\prime }$\ is distant from the origin $\frak{r}$\
approaches zero. This means that the three brane becomes bigger
for $\theta _{\max }^{\prime }$ corresponding to the maximum
values of $\frak{r}_{\max }$\ and shrinks to zero size otherwise.
This result agree perfectly with the fact that the shape of the
wrapped brane depend in the strings attached on it and exercising
forces to keep it inflated. In our cases, there is at the maximum
one string
 attached on the three brane and thus this brane finishes by shrinking to zero size (as found in
figures \ref {d} and \ref {e}).

\section{Conclusion}

In this paper, we have discussed the decay of the D5 brane wrapped
twice around $\mathbb{R}P^{2}$ giving a three brane in the frame
work of the AdS/CFT correspondence with an orthogonal gauge group.
We have shown that, in the case of the super Yang-Mills theory
with $SU(N)$
gauge group, the baryon vertex is represented by a D5 brane wrapped around $%
S^{5}$ and linked to the boundary by $N$ fundamental strings. For
the case of $SO(N)$ gauge theory, This D5-brane becomes unstable
and transforms to a state of wrapped three brane around a three
cycle in $\mathbb{R}P^{5}$ permitted by the topological
restriction on the discrete torsion. We have studied this decay
qualitatively by computing the energy of the three brane
final state and discuss the behavior of its radius in terms of the angle $%
\theta ^{\prime }$. During this analysis we have made some
assumptions like that all the variables are independent apart the
trial $(t,\frak{r},\theta ^{\prime })$ and the fact that
$\frak{r}$ and $A_{0}$ depend only in $\theta ^{\prime }$ meaning
that we are looking for static solutions. The result we have
gotten agrees with the responsibility of the strings attached on
branes on the form of this later.

\begin{center}
{\huge Acknowledgement}
\end{center}
Many thanks to Stefano Sciuto from the University of Turin for the invitation and the hospitality.
Thanks a lot to all members of the string theory group of Turin University for fruitful discussions
 and especially Marco Blau.
I would like to thank El Hassan Saidi for his scientific and administrative helps and kindness.
I am very grateful to Dr Aziz Rhalami for his help and provision.
¶ ¶

\end{document}